\documentclass{article}

\usepackage{arxiv}

\usepackage[utf8]{inputenc} 
\usepackage[T1]{fontenc}    
\usepackage{hyperref}       
\usepackage{url}            
\usepackage{booktabs}       
\usepackage{amsfonts}       
\usepackage{nicefrac}       
\usepackage{microtype}      
\usepackage{cleveref}       
\usepackage{lipsum}         
\usepackage{graphicx}
\usepackage{doi}

\title{Net-charge particle ratio fluctuations in $pp$ collisions at several LHC energies}

\newif\ifuniqueAffiliation
\uniqueAffiliationtrue

\ifuniqueAffiliation 
\author{ \href{https://orcid.org/0000-0000-0000-0000}{\includegraphics[scale=0.00]{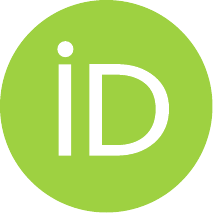}\hspace{1mm}Subhadeep Paul} \\
	Department of Physics\\
	Cooch Behar Panchanan Barma University\\
	Cooch Behar, 736101, India \\
	\And
	\href{https://orcid.org/0000-0000-0000-0000}{\includegraphics[scale=0.00]{orcid.pdf}\hspace{1mm}Tumpa Biswas} \\
	Department of Physics\\
	Cooch Behar Panchanan Barma University\\
	Cooch Behar, 736101, India \\
	\And
	\href{https://orcid.org/0000-0000-0000-0000}{\includegraphics[scale=0.00]{orcid.pdf}\hspace{1mm}Dipak Ghosh} \\
	Sir C.V. Raman Centre for Physics and Music\\
	Jadavpur University\\
	Kolkata, 700032, India\\
	\And
	\href{https://orcid.org/0000-0000-0000-0000}{\includegraphics[scale=0.00]{orcid.pdf}\hspace{1mm}Mehedi Kalam} \\
	Department of Physics\\
	Aliah University\\
	New Town, Kolkata, 700160, India\\
	\And
	\href{https://orcid.org/0000-0002-2765-4544}{\includegraphics[scale=0.06]{orcid.pdf}\hspace{1mm}Prabir Kr. Haldar}\thanks{Corresponding author -- prabirkrhaldar@gmail.com} \\
	Department of Physics\\
	Cooch Behar Panchanan Barma University\\
	Cooch Behar, 736101, India \\
}
\else
\usepackage{authblk}

\newbox{\orcid}\sbox{\orcid}{\includegraphics[scale=0.06]{orcid.pdf}} 
\author[1]{%
	\href{https://orcid.org/0000-0000-0000-0000}{\usebox{\orcid}\hspace{1mm}David S.~Hippocampus\thanks{\texttt{hippo@cs.cranberry-lemon.edu}}}%
}
\author[1,2]{%
	\href{https://orcid.org/0000-0000-0000-0000}{\usebox{\orcid}\hspace{1mm}Elias D.~Striatum\thanks{\texttt{stariate@ee.mount-sheikh.edu}}}%
}
\affil[1]{Department of Computer Science, Cranberry-Lemon University, Pittsburgh, PA 15213}
\affil[2]{Department of Electrical Engineering, Mount-Sheikh University, Santa Narimana, Levand}
\fi

\renewcommand{\shorttitle}{Net-charge particle ratio fluctuations in $pp$ collisions}

\hypersetup{
pdftitle={A template for the arxiv style},
pdfsubject={q-bio.NC, q-bio.QM},
pdfauthor={David S.~Hippocampus, Elias D.~Striatum},
pdfkeywords={First keyword, Second keyword, More},
}

\begin{document}

\maketitle

\begin{abstract}
Event-by-event particle ratio fluctuations for simulated data sets of three different models named UrQMD, AMPT, and Pythia are studied using the fluctuation variable $\nu_{dyn}$. The simulated data sets produced in $pp$ collisions at four different LHC energies $\sqrt{s} = 2.76, 5.02, 7$ and $13$ TeV are generated and considered for this analysis. The variation of fluctuation parameter $\nu_{dyn}$ for accepted pair of meson and baryon combination $i.e.$  $[\pi, K]$, $[\pi, p]$ and $[p, K]$ with the increasing value of the mean multiplicity of charged particles ($\langle N_{ch} \rangle$) are investigated. It has been observed that the correlation between the particle pair $[\pi, K]$ is more than that of the two other particle pairs $[\pi, p]$ and $[p, K]$. However, the energy-wise inspection of the fluctuation variable $\nu_{dyn}$ for $0-10\%$ centrality data shows the increase in the correlation between the particles in each pair for all three models considered. 
\end{abstract}


\section{Introduction}
The experiments involving heavy ions are crafted with the purpose of investigating nuclear material in exceptionally challenging conditions, marked by elevated levels of temperature, density, or both \cite{1}. Within these experiments, there's an intention to generate a novel form of matter termed quark-gluon plasma (QGP). This theoretical state of matter is believed to have influenced the geometry of the cosmic background in the immediate aftermath of the Big Bang, taking shape within a mere few microseconds \cite{2}. This concept finds its roots in the field of quantum chromodynamics (QCD) \cite{3}, and it's theorized to emerge during collisions of heavy ions.

One notable focus is the transition between quarks and hadrons, elemental particles composed of quarks held together by the strong force. This phase transition carries distinctive traits that can be identified through various experimental indicators \cite{4,5,6,7,8,9,10,11}. Among these indicators, the fluctuations in net charge exhibited by the particles produced in these collisions stand out \cite{9,11,12,13}. Correlations and event-by-event (ebe) fluctuations that arise from dynamic processes are believed to be linked to critical phase transition phenomena. Investigating these aspects can reveal local and global distinctions between events that are generated from similar initial conditions \cite{14}.

Researchers have explored fluctuations in ebe data in both hadronic and heavy-ion collisions across a wide range of energies, utilizing various methodologies. These include multifractals \cite{15,16,16a,16b}, normalized factorial moments \cite{17,18,19,20}, k-order rapidity spacing \cite{21,22,23}, erraticity \cite{24,25,26}, and intensive quantities such as those involving multiplicity and transverse momentum \cite{27,28,29}. Additionally, fluctuations in conserved quantities like strangeness, baryon number, and electric charge have emerged as valuable tools to gauge the level of equilibration and criticality in measured systems \cite{12,a,c,d,e}.

Dynamical fluctuations in net charge have been subject to investigation by experiments like the STAR and the ALICE, employing a parameter denoted as $\nu_{dyn}$ \cite{9}. This parameter serves as an effective probe due to its resistance to detector efficiency losses \cite{c}. Alternative measures for net charge fluctuations, such as the variance of charge (V(Q)), variance of charge ratio (V(R)), and the D-measure, are more susceptible to measurement conditions \cite{9,f}. However, there has been a notable observation that emphasizes significant systematic uncertainties within such measurements \cite{8}. These uncertainties often stem from variations in the volume of the system due to changes in impact parameters. Conversely, fluctuations in multiplicity ratios seem to exhibit greater sensitivity to changes in density rather than shifts in volume \cite{7}. Hence, the parameter $\nu_{dyn}$ has been employed as a method to explore the properties of QGP. Unlike conventional definitions based on combinations of similar or dissimilar charges, $\nu_{dyn}$ is determined by considering pairs of particle species \cite{c,i}. The increase and divergence in fluctuation, possibly due to phase transition if occurs, might hold a connection to ebe fluctuations and to illustrate the associated fluctuations in baryon numbers, variations in strangeness, and correlations between baryons and strangeness particle pairs such as $[\pi, K]$, $[\pi, p]$ and $[p, K]$ can be a suitable example \cite{i,j}.

\section{Goal of the study}

Researchers have conducted several studies to examine fluctuations in particle ratios within AA (nucleus-nucleus) collisions. The NA49 experiment studied Pb-Pb collisions in a range of beam energies from 20 to 158 A GeV \cite{k}. The STAR experiment, on the other hand, investigated Au-Au collisions across a center-of-mass (c.m.) energy range from 7.7 to 200 GeV \cite{l}. Additionally, the STAR experiment explored Cu-Cu collisions at c.m. energies $\sqrt{s}=$ 22.4, 62.4, and 200 GeV \cite{m}. These experiments were conducted alongside several others, as referenced \cite{8,i,n}. Particle ratio fluctuations at higher LHC energies have been examined specifically by the ALICE experiment, focusing on collisions at a c.m. energy of 2.76 TeV \cite{o,p,q}.

Particle ratio fluctuations are more pronounced in heavy-ion collisions due to the formation of a quark-gluon plasma and large-scale effects. In proton-proton ($pp$) collisions at LHC energies, these fluctuations are expected to be smaller due to the lack of such collective effects. In $pp$ collisions, the particle production process is governed by simpler mechanisms compared to heavy-ion collisions. While there can still be fluctuations in particle yields due to the inherent probabilistic nature of particle production, these fluctuations are expected to be smaller and less significant than in heavy-ion collisions.

Studying particle ratio fluctuations in $pp$ collisions at LHC energies might still provide valuable insights into the underlying particle production processes and can serve as a baseline for comparison with heavy-ion collisions. However, due to the limitations posed by the smaller system size and absence of a QGP formation, any observed fluctuations in $pp$ collisions would likely be of a different nature and magnitude compared to those seen in heavy-ion collisions.

In the present work, we have studied the correlations and fluctuations among meson-baryon particle pairs $[\pi, K], [\pi, p]$ \& $[p, K]$ using simulated models like UrQMD, AMPT \& Pythia generated data sets at different c.m. energies $\sqrt{s}=$ 2.76, 5.02, 7 \& 13 TeV.  Our main aims of this investigation are: (i) to ensure the existence of correlation among three different particle pairs  $[\pi, K]$, $[\pi, p]$ \& $[p, K]$ using charged particle ratio fluctuation parameter $\nu_{dyn}$ at the mentioned LHC energies, (ii) to study the variation of the strength of correlation between the particles in each pair with the centrality dependent quantity $\langle N_{ch} \rangle$, i.e., the mean charged particle multiplicity, (iii) to find the energy dependence of the values of  $\nu_{dyn}$ of all the three mentioned particle pairs for the most centrality data set (0-10\%), and (iv)  to set a baseline comparison of the data sets generated by the models UrQMD, AMPT, \& Pythia.

\section{Simulated Events Description} 
The Ultra relativistic Quantum Molecular Dynamics \\(UrQMD), A Multi-Phase Transport (AMPT) \& Pythia models are used to simulate $10^5$ MB data events for $pp$ collisions at $\sqrt{s}=$ 2.76, 5.02, 7  \& 13 TeV for the present investigation \cite{31,32,33,34}.
The generated data sets are thus filtered for $\vert \eta \vert < 1.0$ and $0.2 < p_{T}< 5.0$ GeV/$c$ range of pseudorapidity and transverse momentum. The ranges of $\eta$ and $p_{T}$ are selected to work for the region of particle distribution with maximum probability of fluctuations among the produced multiplicities in regime of the ALICE experiment. 

To demonstrate the feasibility of the model-based study, the $p_{T}$ spectra of $\pi, K$ and $p$ produced from UrQMD, AMPT, and Pythia simulated $pp$ interactions with the ALICE experimental data are shown in Fig.~$1$, Fig.~$2$ and Fig.~$3$ respectively. In case of the ALICE data in Fig.~$1$, Fig.~$2$ and Fig.~$3$, data of all the V$0$ classes 
are merged together and their weighted average is said as ALICE minimum bias experimental data in the respective figures \cite{ALICE2020}.

\begin{figure}[h!]
\centering{
\includegraphics[width=.7\columnwidth]{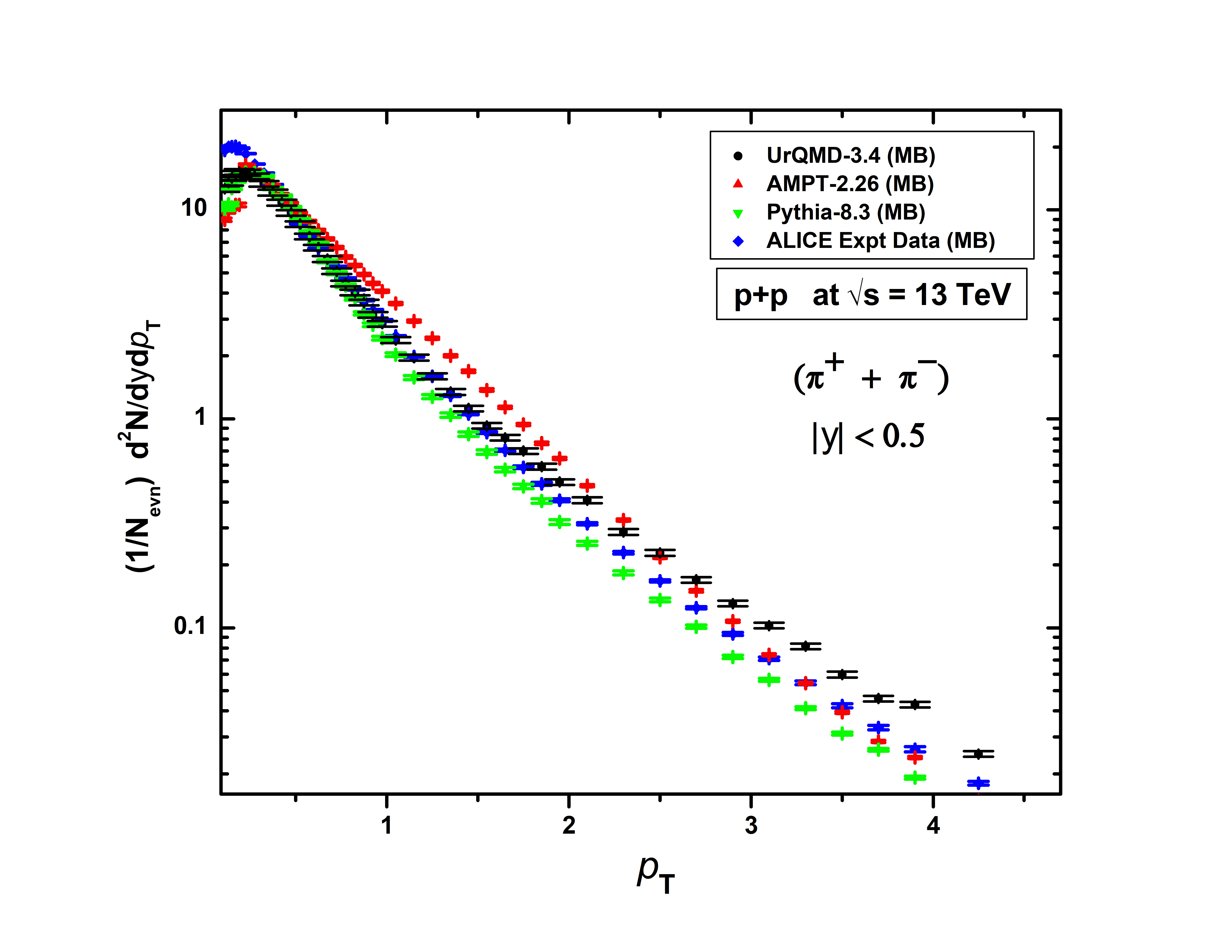}
}
\caption{Comparative analysis of the $p_T$ distribution of pions produced from the different simulated models with the ALICE experimental data \cite{ALICE2020}.}
\end{figure}

\begin{figure}[h!]
\centering{
\includegraphics[width=.7\columnwidth]{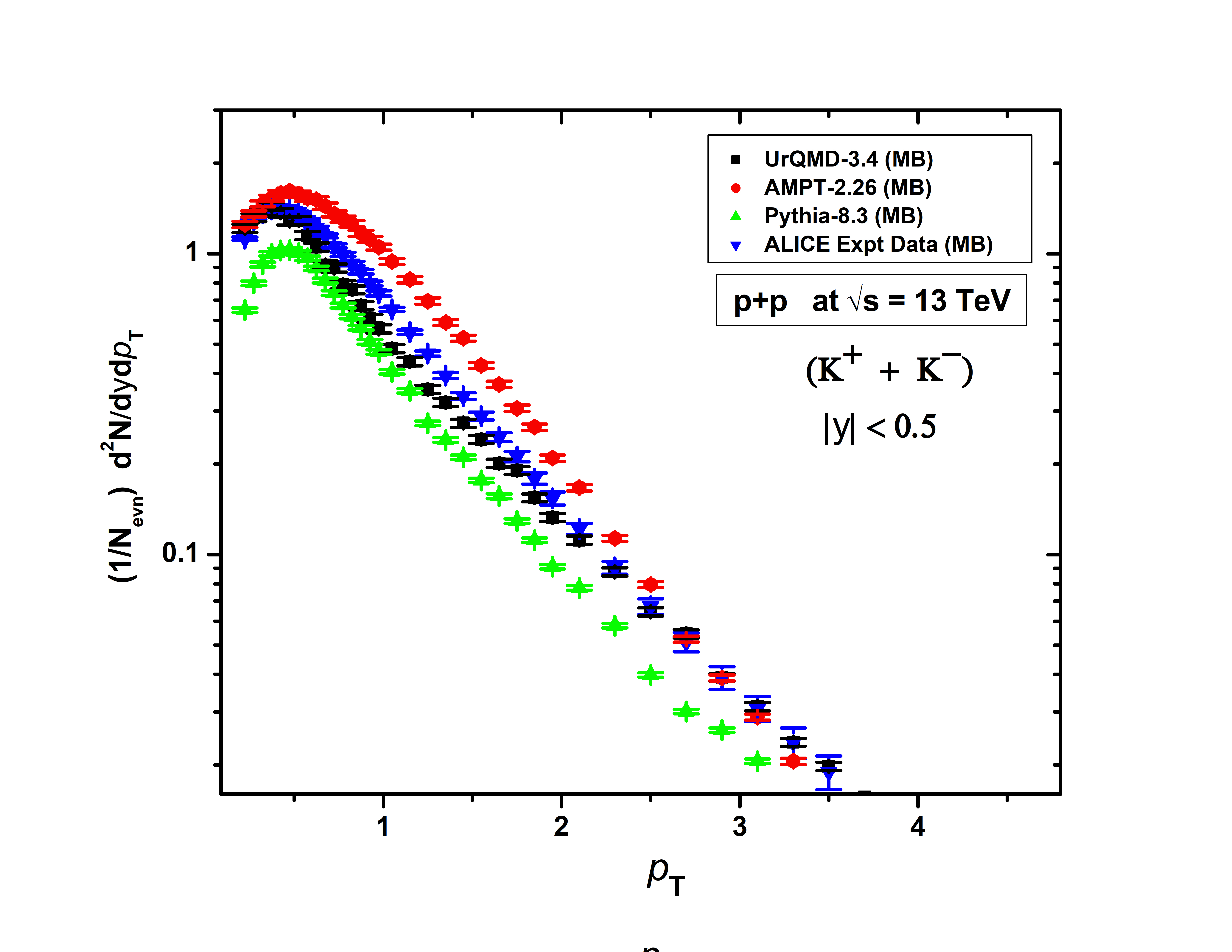}
}
\caption{Comparative analysis of the $p_T$ distribution of kaons produced from the different simulated models with the ALICE experimental data \cite{ALICE2020}.}
\end{figure}

\begin{figure}[h!]
\centering{
\includegraphics[width=.7\columnwidth]{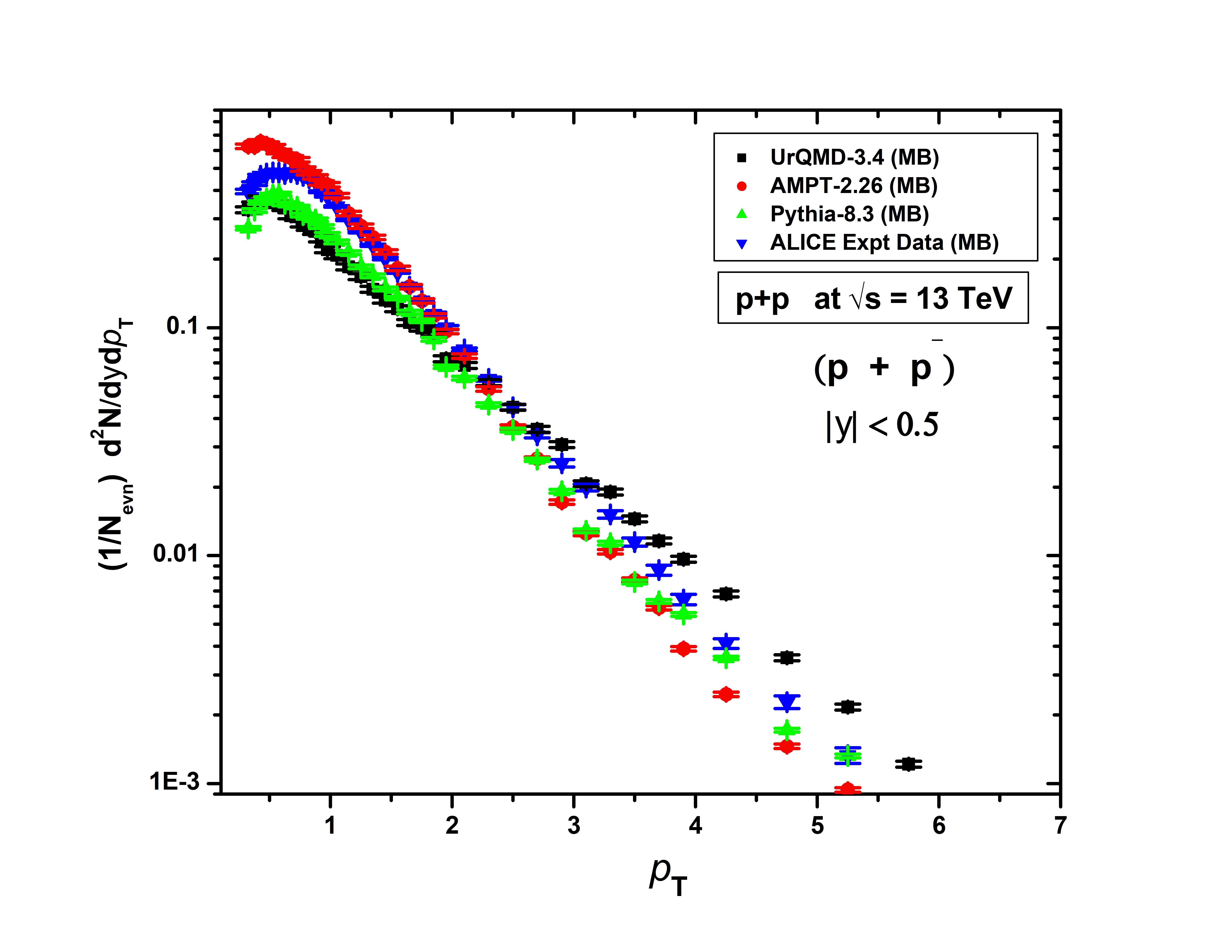}
}
\caption{Comparative analysis of the $p_T$ distribution of protons and anti-protons produced from the different simulated models with the ALICE experimental data \cite{ALICE2020}.}
\end{figure}

\subsection{UrQMD v3.4}
  To explore the known and speculative mechanisms responsible for generating multiplicities of charged particles, the researchers have employed the UrQMD-3.4 model for simulation purposes \cite{35,36,37}. This microscopic transport model, UrQMD, is focused on the covariant propagation of color strings and incorporates degrees of freedom related to mesons and baryons. Within this framework, the model accounts for the creation of hadronic resonances, particle resonances, and the fragmentation of color strings. Notably, these characteristics are commonly observed in experiments conducted at RHIC and LHC energy levels. For the lower energy range, approximately 5.0 GeV, the UrQMD model provides an understanding of collisions between atomic nuclei (referred to as AB collisions), contingent upon the influence of hadronic degrees of freedom. Collisions between two particles become apparent when the impact parameter $ b<\sqrt{\frac{\sigma_t}{\pi}}$ \cite{38}, here $\sigma_t$ denotes the total cross section. The UrQMD model encompasses the AB interaction, which is an accumulation of nucleon-nucleon (NN) collisions. This property renders the UrQMD model suitable for researchers to investigate AB interactions within the energy range relevant to their requirements. However, it's important to note that certain significant phenomena, such as the transition between quark-gluon plasma (QGP) and hadron gas (HG) phases, are not explicitly incorporated within the UrQMD model. Instead, the model adheres to Hagedron-type dynamics \cite{39}. It describes the intermediate fireball using a local thermal and chemical equilibrium framework. The presence of Bose-Einstein correlations is likely to influence fluctuations, which are taken into account for measurement purposes. This accounts for the existence of short-range correlations among the multiplicities of charged particles in high-energy interactions \cite{40}. In the UrQMD-3.4 version, a more accurate iso-energy density particlization hypersurface has been integrated \cite{41}.  
  
\subsection{AMPT v2.26}
The AMPT (A Multi-Phase Transport) model is composed of four essential components: initial conditions, partonic interactions, the transition from partonic to ha-dronic matter, and hadronic interactions \cite{42}. The initial conditions are derived from the HIJING model and encompass the spatial and momentum distributions of minijet partons and soft string excitations. Zhang's parton cascade (ZPC) is responsible for simulating parton scatterings, focusing exclusively on two-body scatterings with cross sections obtained from perturbative QCD (pQCD) incorporating screening masses.  In the default AMPT model, partons are combined with their parent strings once their interactions cease, and these resulting strings are then fragmented into hadrons using the Lund string fragmentation model \cite{43,44,45,46,47,48}. However, in the AMPT model employing the string melting condition, quark coalescence is employed instead of recombining partons into hadrons. Subsequently, a hadronic cascade, initially based on the ART model, is expanded to include additional reaction pathways that become significant at higher energies. This cascade process is employed to describe the dynamics of the subsequent hadronic phase. These channels include the creation of baryons and antibaryons from mesons, along with the reverse reactions of annihilation, as well as the formation and decay of $K^*$ resonance and antibaryon resonances. 

  The AMPT model incorporates the concept of string melting. In this mechanism, excited strings that are not part of the incoming projectiles and have not directly interacted with nucleons are converted into partons based on the flavors and spin orientations of their valence quarks. When a meson transforms, specifically into a quark and an anti-quark, or when a baryon transforms, it follows a two-step process. Initially, a baryon is transformed into a quark and a diquark. This transformation is guided by weights determined from relationships found in the SU(6) quark model. Subsequently, the diquark splits into two individual quarks. It is assumed that the quark and diquark masses are the same as in the Pythia programme , i.e. $m_u$ = 5.6 MeV/$c^2$, $m_d$ = 9.9 MeV/$c^2$, and $m_s$ = 199 MeV/$c^2$. The production period of the partons is given by $t_f$ = $E_H$ /$m^2 _{T,H}$ , where $E_H$ and $m _{T,H}$ are the energy and transverse mass of the parent hadron respectively. The process of two-body decomposition described above occurs uniformly in the rest frame of the parent hadron or diquark. When the AMPT model incorporates string melting, its results align with those of the HIJING model when there are no interactions between partons and hadrons. This is because, in the absence of interactions, the partons would naturally approach each other as the closest partners at the same freeze-out time, effectively recombining into the original hadrons.

In the AMPT string melting model, interactions between quarks and antiquarks of various types are taken into account. In the context of the string melting scenario, the straightforward quark coalescence model is employed to simulate the hadronization of partons once their interactions have ceased. In this model, the two closest partons are merged to form a meson, while the three nearest quarks (or antiquarks) are combined to create a baryon (or antibaryon).

When partons combine to form a hadron, preserving a well-defined 4-momentum becomes challenging due to the fact that the invariant mass creates a continuous spectrum rather than a discrete one. The three-momentum is now preserved throughout coalescence, and the flavour and invariant mass of coalescing partons are used to identify the hadron species. The produced meson  has a mass that is closer to the invariant mass of the coalescing quark and antiquark pair for pseudo-scalar and vector mesons with the same flavour composition. The aforementioned quark coalescence model incorporates mesons and baryons which are enlisted in the HIJING program, excluding $\eta$, $\Xi^*$ and $\Sigma^*$ which are not included in hadronic transport model. The process of hadronization leads to the emergence of a phase in which both partons and hadrons coexist. This coexistence arises because partons dynamically freeze out at different stages within the parton cascade. Consequently, the creation of hadrons through their coalescence occurs at various points in time. Moreover, the interactions between these particles account for the inclusion of numerous higher resonances, which act as intermediate states. Within the AMPT model, the progression of hadrons is described using the ART (A Relativistic Transport) model. Originally designed for heavy ion collisions at AGS (Alternating Gradient Synchrotron) energy levels, the ART model forms the basis of the hadron cascade in the AMPT framework \cite{49}.
 
\subsection{Pythia v8.3}
Pythia addresses a wide array of phenomenological challenges in particle physics, as well as related topics in astro-particle, nuclear, and neutrino physics. Its foundation lies in the Lund string model of hadronization. This model is most suitable when dealing with hadronizing systems with invariant masses equal to or exceeding 10 GeV, but its reliability diminishes for systems with lower masses. For situations involving lower-mass systems, typically integrated into larger events, such as those involving color reconnections, hadronic rescattering, or heavy-flavour decays, the model is still used. For the very lowest-mass systems emitting only a couple of hadrons, a simplified cluster-style model called "ministrings" is employed, while standard string fragmentation is applied for other cases.

However, Pythia offers more than just string hadron-ization. It encompasses advanced simulations for various particle-physics events. The structure of the Pythia v8.3 Monte-Carlo event generator is designed to accommodate the diverse physics descriptions and models necessary to create entirely exclusive final states as observed in collider experiments. The layout of Pythia v8.3 is divided into three key segments: process level, parton level, and hadron level. These divisions collectively enable the simulation of comprehensive and distinct final states, mirroring what is observed in real collider experiments \cite{50}. The hard-scattering process, which produces transient resonances, is represented at the process level.  In most cases, the hard process is represented perturbatively with a small amount of particles, often at high-energy ranges \cite{51}. At the parton level in Pythia, there are several shower models available that account for initial- and final-state radiation. This stage includes the consideration of multiparton interactions, processing of beam remnants, and the probability of color-reconnection phenomena. The outcome of this parton-level development displays the actual partonic structure, including jets and a description of the underlying event. The confinement of partons according to QCD principles into color-singlet systems takes place at the hadron level.

In Pythia v8.3, hadronization is depicted as QCD strings breaking apart to form hadrons. Unstable hadrons subsequently decay, and the rescattering of hadrons is accounted for at the hadron level. Since hadronization physics models often involve non-perturbative aspects, they require modeling and parameter adjustment. The output at the hadron level represents an event as it would be detected in an actual experiment. The interface between parton showers and the process level employs a matching and merging mechanism. Parton distribution functions play a crucial role at both the process level and during Initial State Radiation (ISR). The "Info" object is utilized to store and retrieve essential data across all levels of the simulation.

In the analysis of heavy-ion collisions using Pythia v8.3, various parton-level objects can be employed to represent independent sub-collisions. These sub-collisions can then be merged to produce the final hadronization outcome.

\section{Method of Analysis}
 The net-charge fluctuations of the produced particles in high energy nucleus-nucleus collisions is one of the different signatures through which phase-transition and formation of QGP can be studied. $\nu_{dyn}$ parameter is commonly accepted for investigating the particle ratio fluctuations in terms of the ratio of yeild particle species $A$ and $B$ \cite{9,f}.
 
 The standard deviation which describes the fluctuations in $A/B$ ratio is given by the following equation \cite{63,64,65} 
 
\begin{equation}
\label{eq:first}
\sigma^{2}_{A/B} = \frac{\langle N^2_{A} \rangle - \langle N_{A} \rangle^{2}}{\langle N_A \rangle^2} + \frac{\langle N^2_{B} \rangle - \langle N_{B} \rangle^{2}}{\langle N_B \rangle^2} \\ -  2 \frac{\langle N_A N_B \rangle - \langle N_A \rangle \langle N_B \rangle}{\langle N_A \rangle \langle N_B \rangle}
\end{equation} 
where $N_A$ and $N_B$, respectively,  represents the event multiplicities of particle type $A$ and $B$ within the given kinematical limits, while the quantities within $\langle ... \rangle$ represent their mean values. It is worthwhile to mention that the particle type $A$ and $B$ includes the particle and its anti-particle. The covariance measures how much two random variables change with each other. For identical variables, the variance is the special case of covariance. The quantity $\langle N_{A} N_{B} \rangle$ gives the statistical average of the integrals or multiplicities of the simultaneous correlation of the production of both $A$ and $B$ type particles. 

The equation (\ref{eq:first}) can be rewritten as

\begin{equation}
\label{eq:second}
\sigma^{2}_{A/B} = \frac{\langle N_A (N_A - 1) \rangle}{\langle N_A \rangle ^2} + \frac{\langle N_B (N_B - 1) \rangle}{\langle N_B \rangle ^2} \\ - 2 \frac{\langle N_A N_B\rangle}{\langle N_A \rangle \langle N_B \rangle}  + \frac{1}{\langle N_A \rangle} + \frac{1}{\langle N_B \rangle}
\end{equation}

The last two terms in equation (\ref{eq:second}) represent uncorrelated particle production which is statistical in nature in Poisson's limit. 
The charge dependence of the dynamical net-charge fluctuations for $A/B$ charge ratio is given by $\nu_{dyn}[A,B]$ which measures the deviation of the fluctuations in the multiplicity of particle species $A$ and $B$ from that expected from Poissonian statistics \cite{5}. The first three terms in equation (\ref{eq:second}) is better described by $\sigma^2_{dyn}$. $\nu_{dyn}[A,B]$ variable does not involve particle ratios directly but it is related to $\sigma_{dyn}$ as $\sigma^2_{dyn} \approx \nu_{dyn}[A,B]$ \cite{5,m}. 
Thus, $\nu_{dyn}[A,B]$ is given as \cite{5,i,o,p,q}

\begin{equation}
\label{eq:third}
\nu_{dyn}[A,B]= \frac{\langle N_A (N_A - 1) \rangle}{\langle N_A \rangle ^2} + \frac{\langle N_B (N_B - 1) \rangle}{\langle N_B \rangle ^2} \\ - 2 \frac{\langle N_A N_B\rangle}{\langle N_A \rangle \langle N_B \rangle}
\end{equation}

If particles $A$ and $B$ are produced in statistically independent way then $\nu_{dyn}[A,B]$ should be zero \cite{9,o,64}. However, as the particles produced in the collision process are partially correlated through the production of resonances, string fragmentation, jet fragmentation, and (or) other mechanisms, the value of $\nu_{dyn}[A,B]$ is expected to be a nonzero value \cite{5}. Dominance of first two terms in equation (\ref{eq:third}) gives positive value of $\nu_{dyn}[A,B]$ which indicates the presence of anticorrelation whereas the negative value of $\nu_{dyn}[A,B]$ is due to the dominance of the third term in equation (\ref{eq:third}) which indicates the correlation between $A$ and $B$ type of particles. 
In this paper we have considered three pair of particle's combination such as $[\pi, K]$, $[\pi, p]$ and $[p, K]$ for the indices $A$, $B$.

Subsample method is used to determine the statistical errors associated to $\nu_{dyn}$ \cite{k}. Subsamples made out of data set are used to calculate the values of $\nu_{dyn}$, which are then used to estimate the mean and dispersion as:

\begin{equation}
\label{eq:fourth}
< \nu_{dyn}[A,B] > = \frac{1}{n} \Sigma \nu_{dyn}[A,B]_{i},
\end{equation}

\begin{equation}
\label{eq:fifth}
\sigma_{dyn} = \sqrt{\frac{\Sigma (\nu_{dyn}[A,B]_{i} - <\nu_{dyn}[A,B]>)^2}{n-1}}
\end{equation}

Therefore, the associated statistical error is calculated as

\begin{equation}
\label{eq:sixth}
(Error)_{stat} = \frac{\sigma_{dyn}}{\sqrt{n}}
\end{equation}

\section{Result and Discussions}
In this investigation of event-by-event particle ratio fluctuations using $\nu_{dyn}$ variable, simulated minimum bias (MB) events are generated using Pythia v8.3, AMPT v2.26 and UrQMD v3.4 for $pp$ collisions at $\sqrt{s}= 2.76,$ $5.02, 7$ and $13$ TeV. The number of events generated for each of this model is $10^5$. Charged particles having pseudorapidity ($\eta$) and transverse momentum ($pT$) in the range $\vert \eta \vert < 1.0$ and $0.2 < pT < 5.0 GeV/c$ in regime of the ALICE experiment is considered out of the MB data set for this analysis. Unlike the centrality estimation of the ALICE experiment, for the purpose of simplicity in this analysis, six centrality classes ranging from $0-60\%$ at an interval of $10\%$ is made depending on higher multiplicities as most central data set ($0-10\%$) and the successive data sets as in accordance.  

Variation of $\nu_{dyn}$ with the mean multiplicity of charged particles ($\langle N_{ch} \rangle$) calculated from each centrality data set is represented in figs.~$4-15$ for accepted pair of meson and baryon combination $i.e.$  $[\pi, K]$, $[\pi, p]$ and $[p, K]$. \\
\begin{figure}[h!]
\centering{
\includegraphics[width=.7\columnwidth]{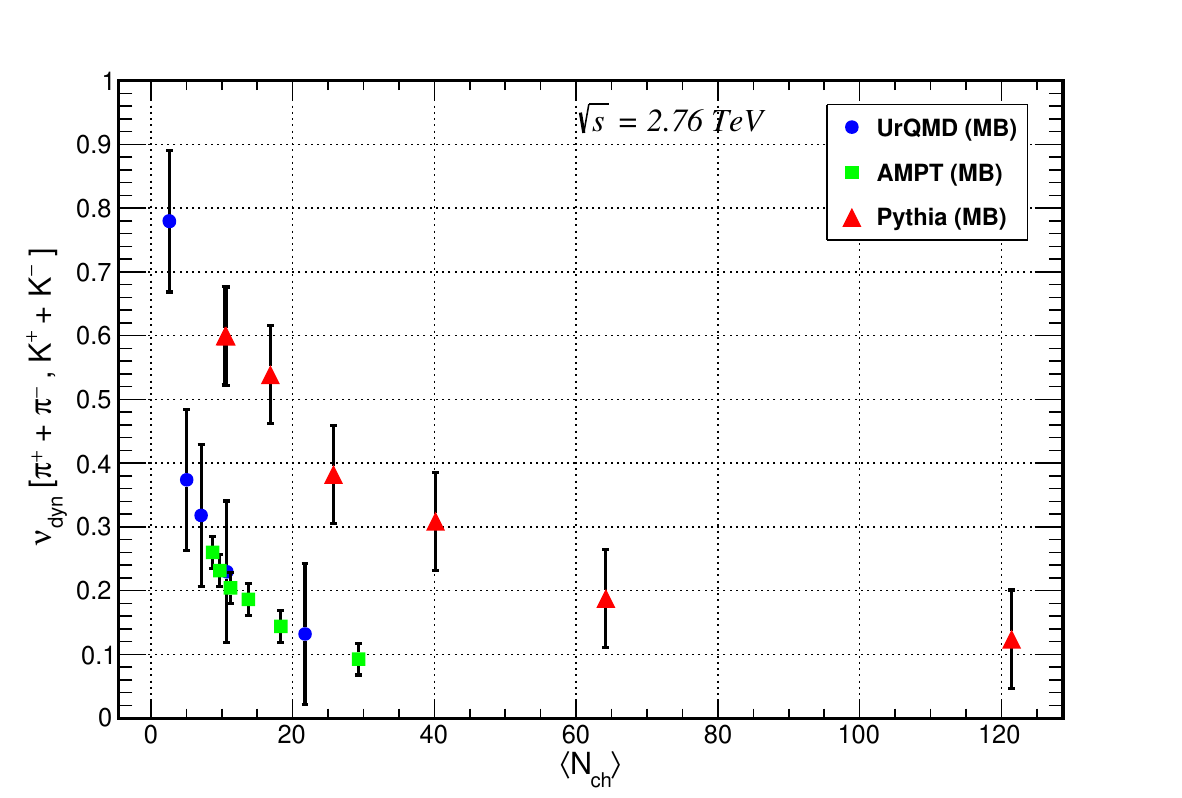}
}
\caption{The variation of $\nu_{dyn}[\pi, K]$ with mean charged particle multiplicity $\langle N_{ch} \rangle$ at $\sqrt{s} = 2.76$ TeV.}
\end{figure}

\begin{figure}[h!]
\centering{
\includegraphics[width=.7\columnwidth]{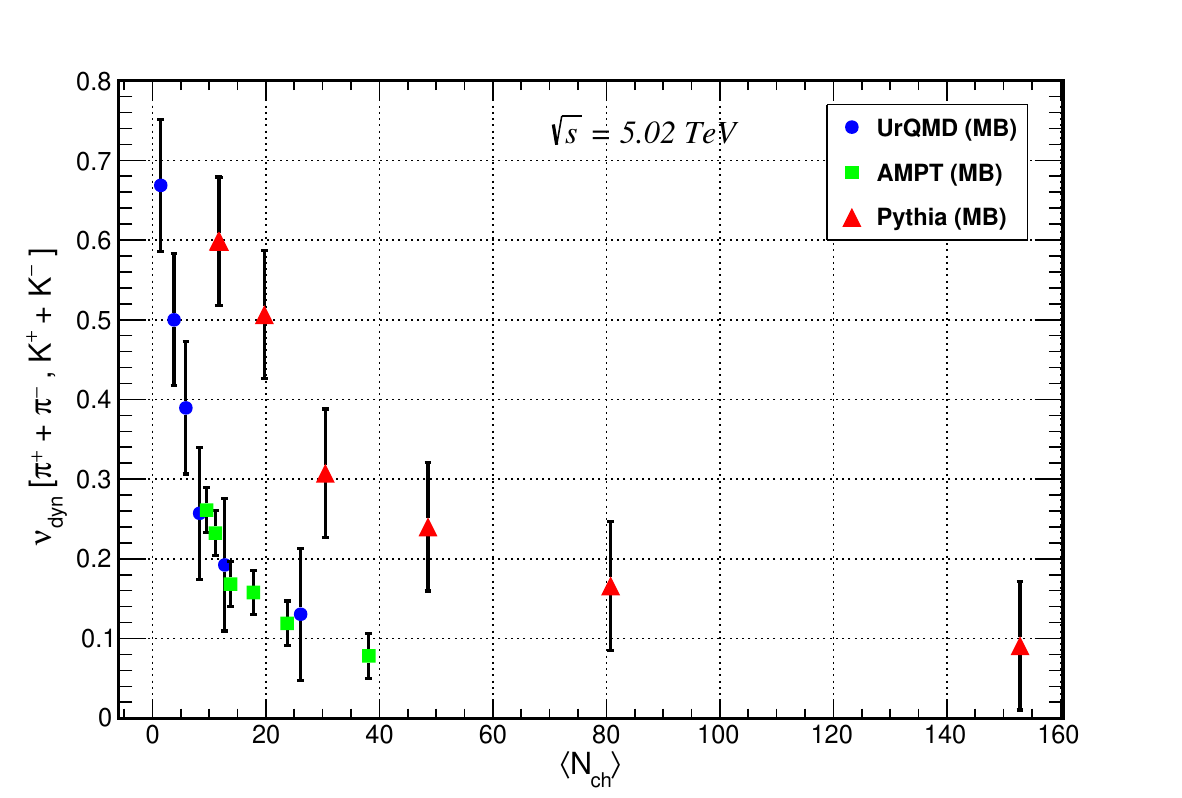}
}
\caption{The variation of $\nu_{dyn}[\pi, K]$ with mean charged particle multiplicity $\langle N_{ch} \rangle$ at $\sqrt{s} = 5.02$ TeV.}
\end{figure}

\begin{figure}[h!]
\centering{
\includegraphics[width=.7\columnwidth]{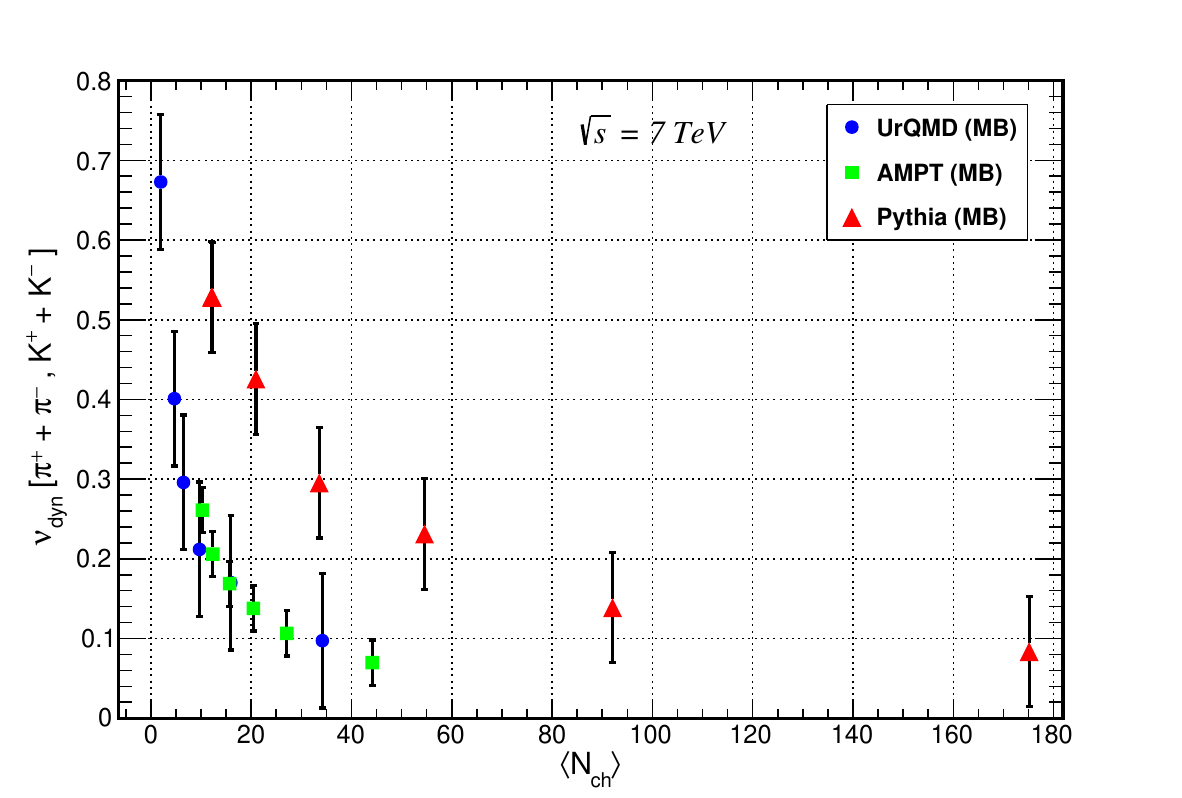}
}
\caption{The variation of $\nu_{dyn}[\pi, K]$ with mean charged particle multiplicity $\langle N_{ch} \rangle$ at $\sqrt{s} = 7$ TeV.}
\end{figure}

\begin{figure}[h!]
\centering{
\includegraphics[width=.7\columnwidth]{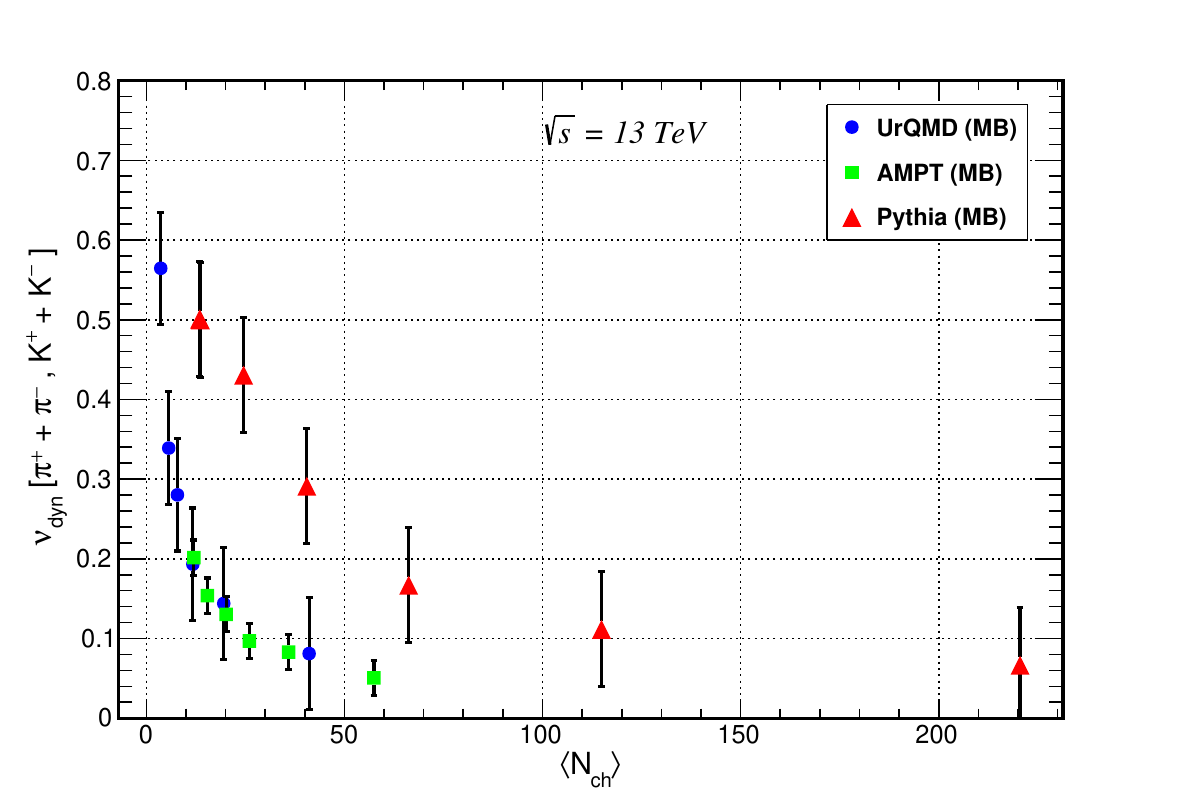}
}
\caption{The variation of $\nu_{dyn}[\pi, K]$ with mean charged particle multiplicity $\langle N_{ch} \rangle$ at $\sqrt{s} = 13$ TeV.}
\end{figure}

Evidently, quark number fluctuation and correlation influences the variance and two-hadron correlations and therefore $\nu_{dyn}[\pi, K]$ \cite{73,74}. First four figures, figs.~$4-7$ shows the $\pi-K$ fluctuations $\nu_{dyn}[\pi, K]$ respectively at four different LHC energy, $\sqrt{s}= 2.76, 5.02, 7$ and $13$ TeV, where $\pi$ refers to $\pi^+ + \pi^-$ and $K$ refers to $K^+ + K^-$. It may be noted that the value of $\nu_{dyn}$ is maximum for the smallest value of $\langle N_{ch} \rangle$, $i.e.$ for peripheral collisions and it decays quickly for larger values of $\langle N_{ch} \rangle$ at respective energies for all the three data sets of Pythia, AMPT and UrQMD. The positive values of $\nu_{dyn}$ indicates the presence of rather stronger anticorrelation in peripheral collisions with low $\langle N_{ch} \rangle$. 
All the three data-sets of UrQMD, AMPT and Pythia shows similar falling variation but the UrQMD data repesents higher value of $\nu_{dyn}$ $i.e.$ the fluctuation between produced $\pi$ \& $K$ is more in UrQMD data in comparison to Pythia and AMPT data. The AMPT data has least fluctuations measured between $\pi$ \& $K$ at all the four concerned energies in this discussion. It is also to note that the larger value of $\langle N_{ch} \rangle$ shows the production of charged particles in Pythia is much higher than the other two models UrQMD and AMPT. The value of $\langle N_{ch} \rangle$ corresponds to each centrality data sets increases with the increase in the LHC energy from $\sqrt{s}= 2.76$ TeV to $\sqrt{s}= 13$ TeV as increase in centre-of-mass (c.m.) energy gives rise to  more number of particle production in all the three models studied. 

\begin{figure}[h!]
\centering{
\includegraphics[width=.7\columnwidth]{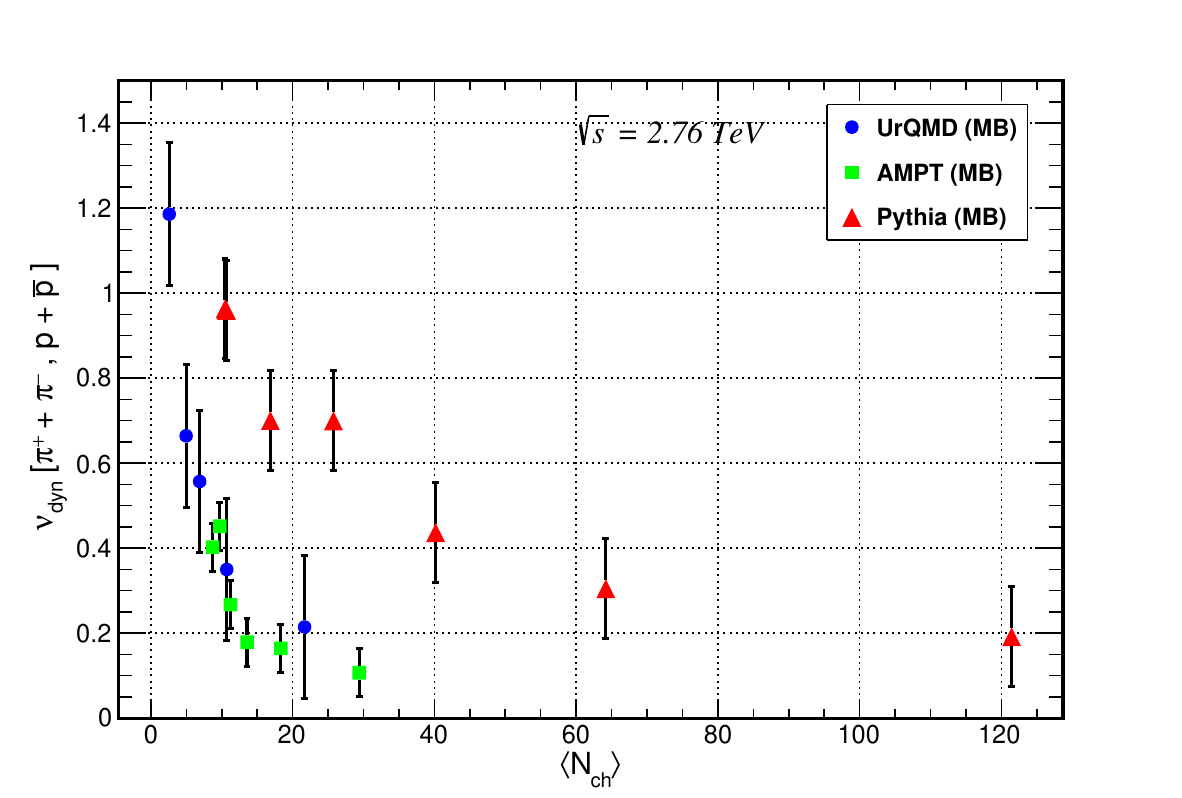}
}
\caption{The variation of $\nu_{dyn}[\pi, p]$ with mean charged particle multiplicity $\langle N_{ch} \rangle$ at $\sqrt{s} = 2.76$ TeV.}
\end{figure}

\begin{figure}[h!]
\centering{
\includegraphics[width=.7\columnwidth]{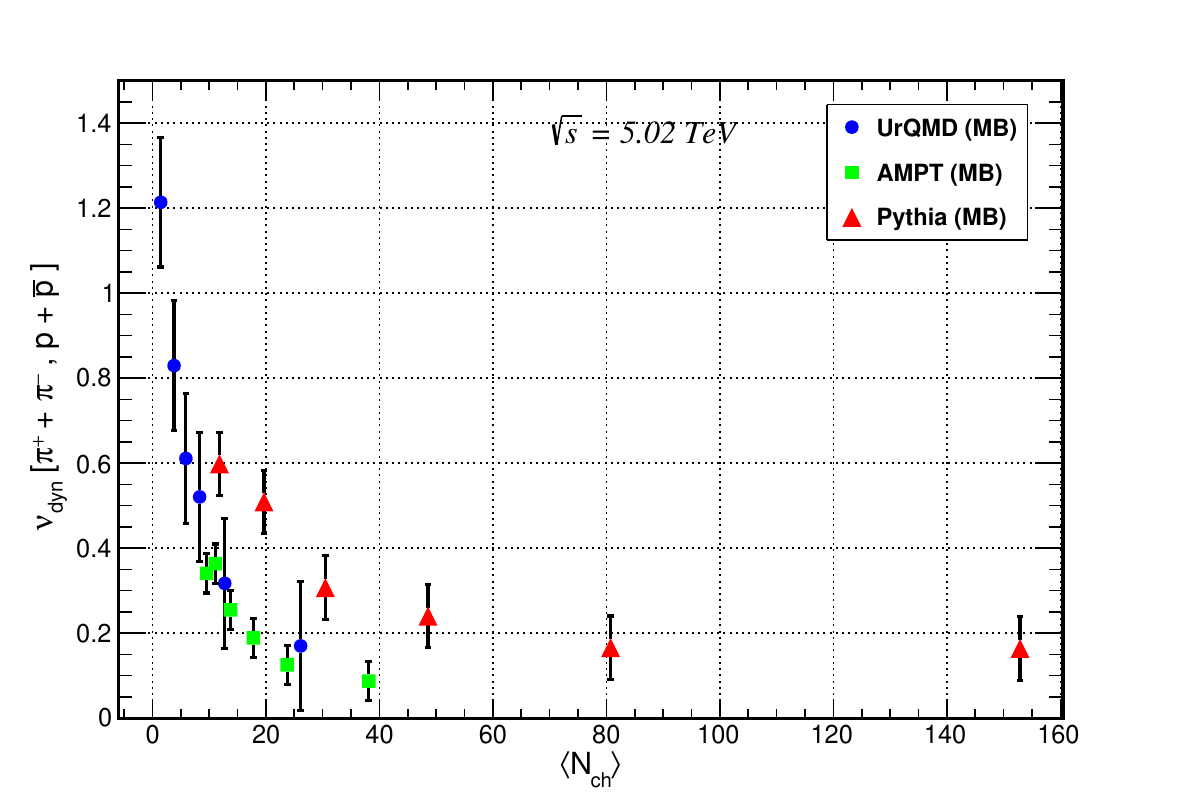}
}
\caption{The variation of $\nu_{dyn}[\pi, p]$ with mean charged particle multiplicity $\langle N_{ch} \rangle$ at $\sqrt{s} = 5.02$ TeV.}
\end{figure}

\begin{figure}[h!]
\centering{
\includegraphics[width=.7\columnwidth]{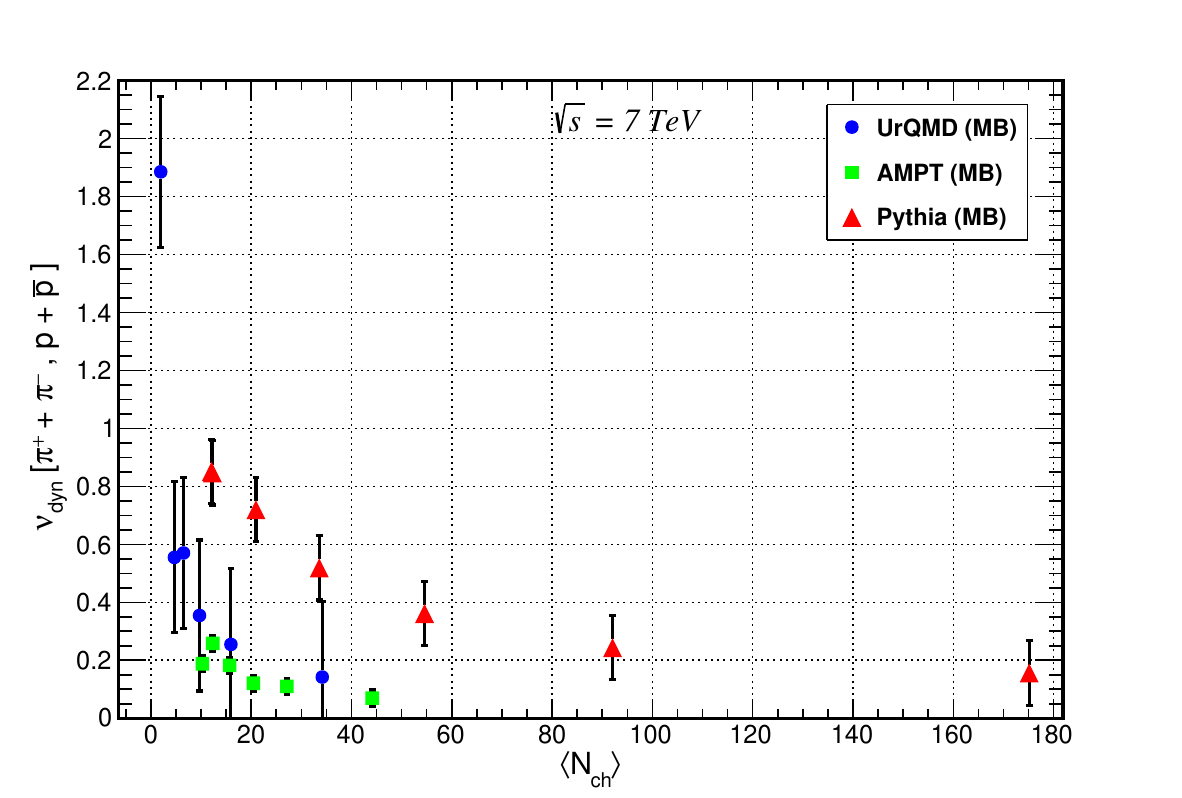}
}
\caption{The variation of $\nu_{dyn}[\pi, p]$ with mean charged particle multiplicity $\langle N_{ch} \rangle$ at $\sqrt{s} = 7$ TeV.}
\end{figure}

\begin{figure}[h!]
\centering{
\includegraphics[width=.7\columnwidth]{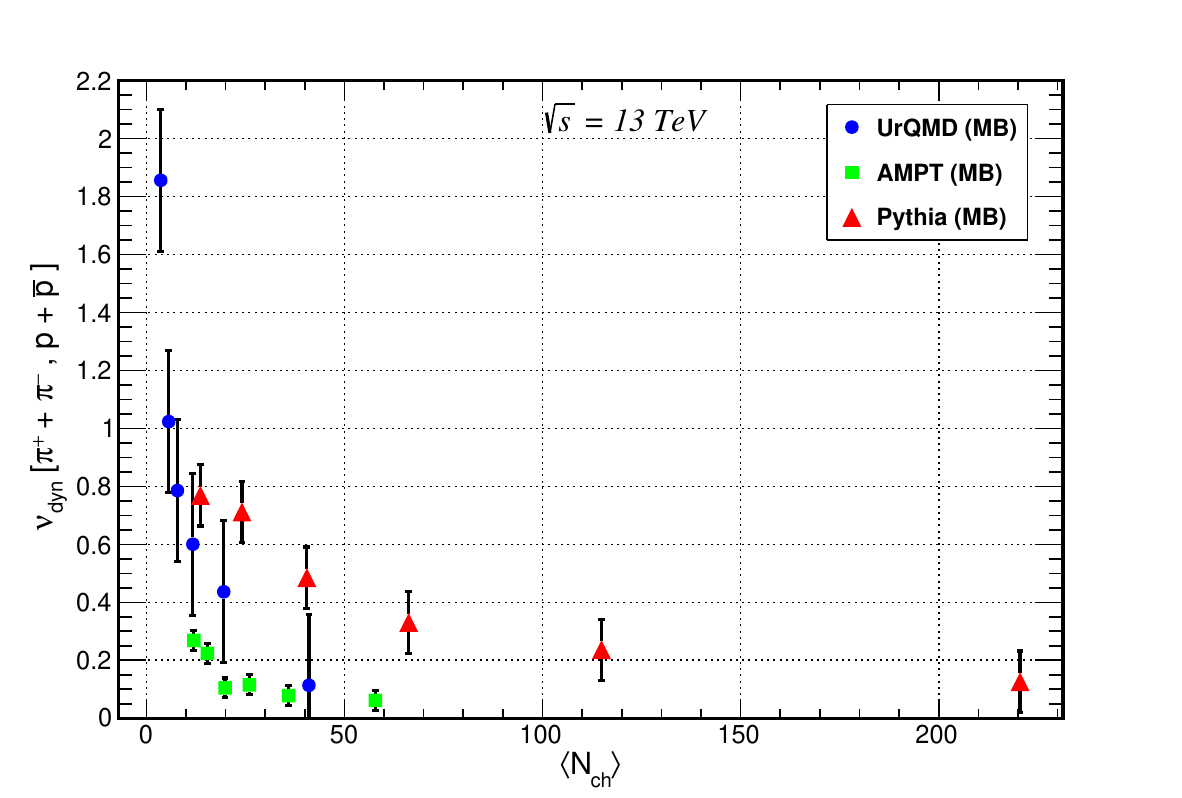}
}
\caption{The variation of $\nu_{dyn}[\pi, p]$ with mean charged particle multiplicity $\langle N_{ch} \rangle$ at $\sqrt{s} = 13$ TeV.}
\end{figure}

The particle ratio fluctuation bewteen $\pi$ and $p$ with $\langle N_{ch} \rangle$ is studied and shown in figs.~$8-11$ for $\sqrt{s}= 2.76, 5.02, 7$ and $13$ TeV respectively, where $\pi$ refers to $\pi^+ + \pi^-$ and $p$ refers to $p + \overline{p}$. The variation of all the data sets in this case is similar to the variation of $\nu_{dyn}[\pi, K]$ with $\langle N_{ch} \rangle$, $i.e.$ the value of $\nu_{dyn}[\pi, p]$ is maximum for lower value of $\langle N_{ch} \rangle$ and then decreases as $\langle N_{ch} \rangle$ increases for most central collisions. It is very important to note that the $\pi$ and $p$ fluctuation is little more pronounced compared to the fluctuation between  $\pi$ and $K$ for all the generated data sets of Pythia, AMPT and UrQMD at mentioned energies. 
At a certain energy, the inspection of $\pi$ and $p$ fluctuations with $\langle N_{ch} \rangle$ shows, the UrQMD generated data sets has maximum $\pi$ and $p$ fluctuations and AMPT has least fluctuations in between $\pi$ and $p$ among all the data sets generated using Pythia, AMPT and UrQMD. Figs.~$8-11$ also shows, the fluctuations between $\pi$ and $p$ in AMPT data especially at $\sqrt{s}= 2.76, 5.02$ and $7$ TeV has a disturbance in the falling trend of $\nu_{dyn}[\pi, p]$ with increasing $\langle N_{ch} \rangle$ certainly at lower value of $\langle N_{ch} \rangle$, perhaps the statistical fluctuation is being the reason behind this. 

\begin{figure}[h!]
\centering{
\includegraphics[width=.7\columnwidth]{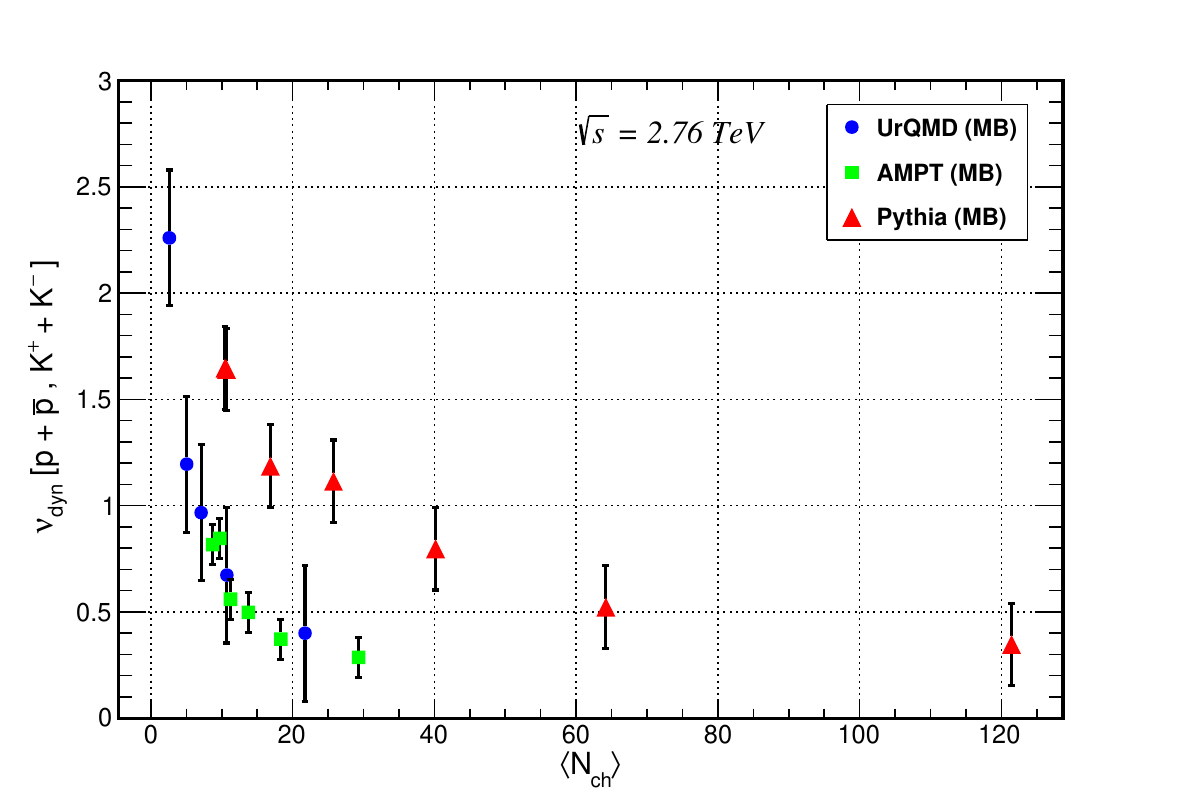}
}
\caption{The variation of $\nu_{dyn}[p, K]$ with mean charged particle multiplicity $\langle N_{ch} \rangle$ at $\sqrt{s} = 2.76$ TeV.}
\end{figure}

\begin{figure}[h!]
\centering{
\includegraphics[width=.7\columnwidth]{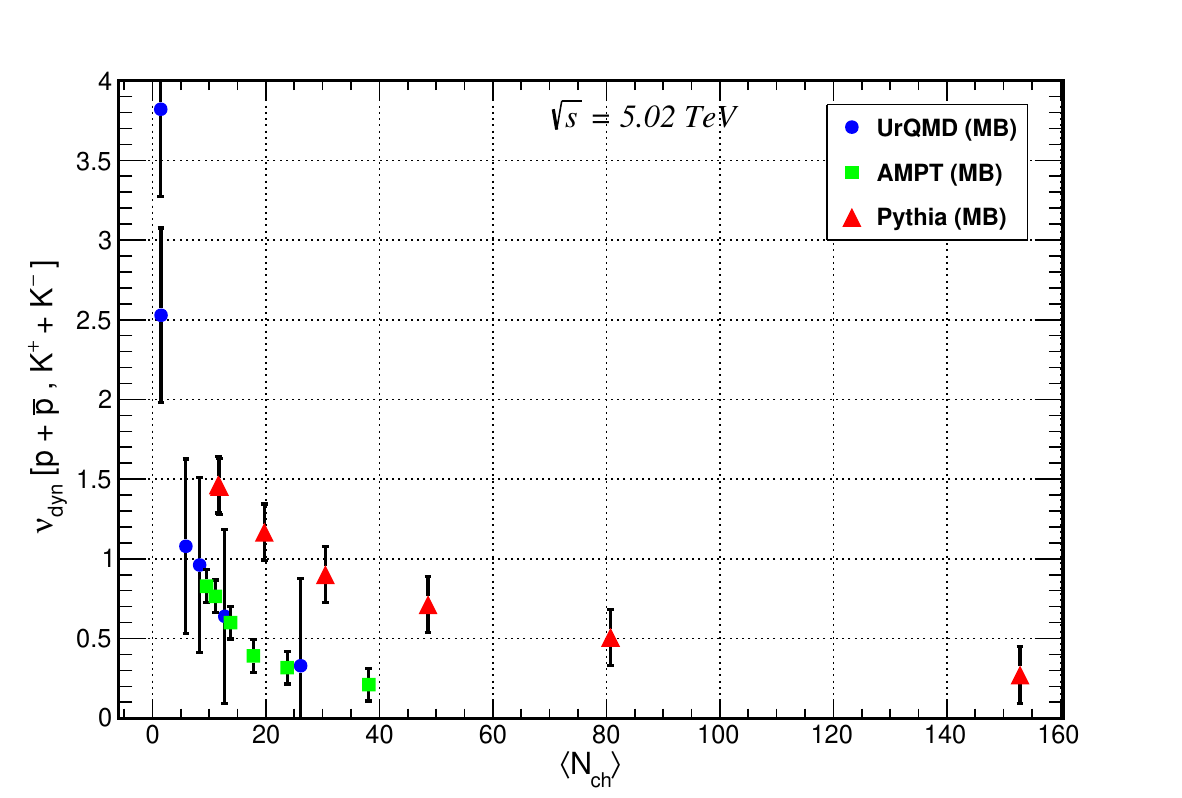}
}
\caption{The variation of $\nu_{dyn}[p, K]$ with mean charged particle multiplicity $\langle N_{ch} \rangle$ at $\sqrt{s} = 5.02$ TeV.}
\end{figure}

\begin{figure}[h!]
\centering{
\includegraphics[width=.7\columnwidth]{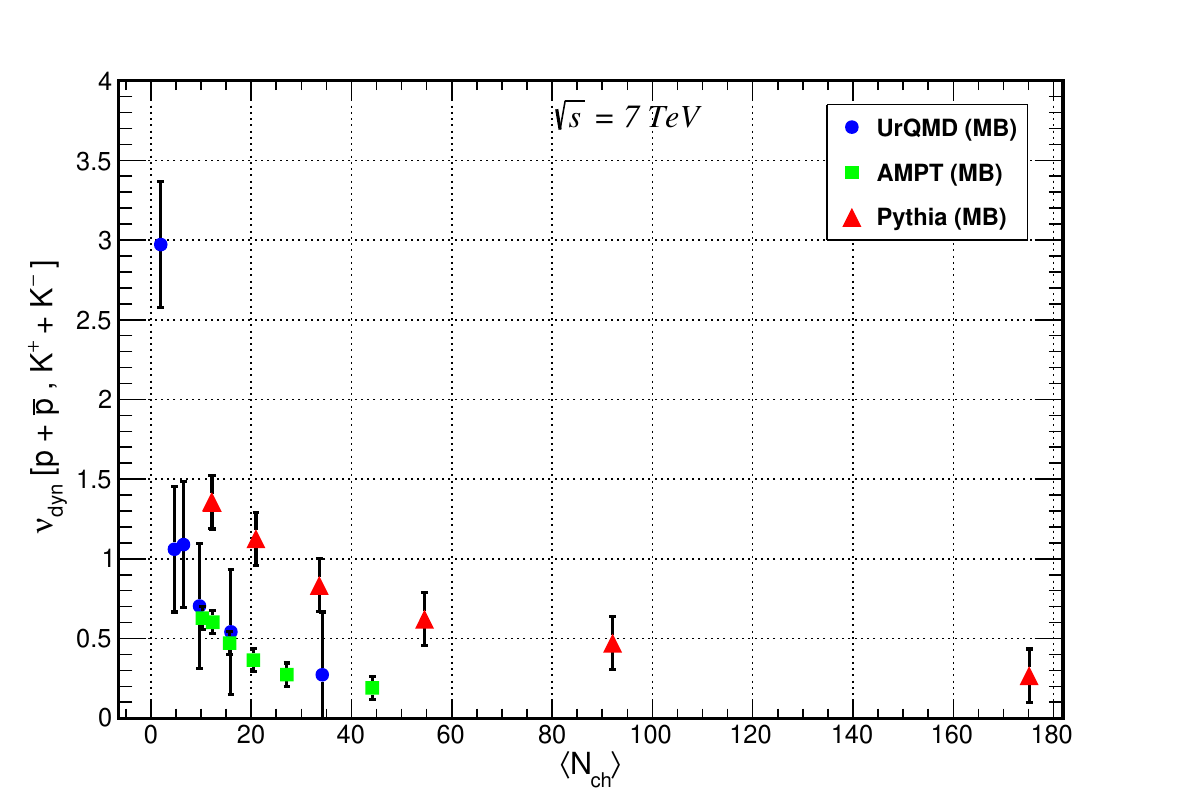}
}
\caption{The variation of $\nu_{dyn}[p, K]$ with mean charged particle multiplicity $\langle N_{ch} \rangle$ at $\sqrt{s} = 7$ TeV.}
\end{figure}

\begin{figure}[h!]
\centering{
\includegraphics[width=.7\columnwidth]{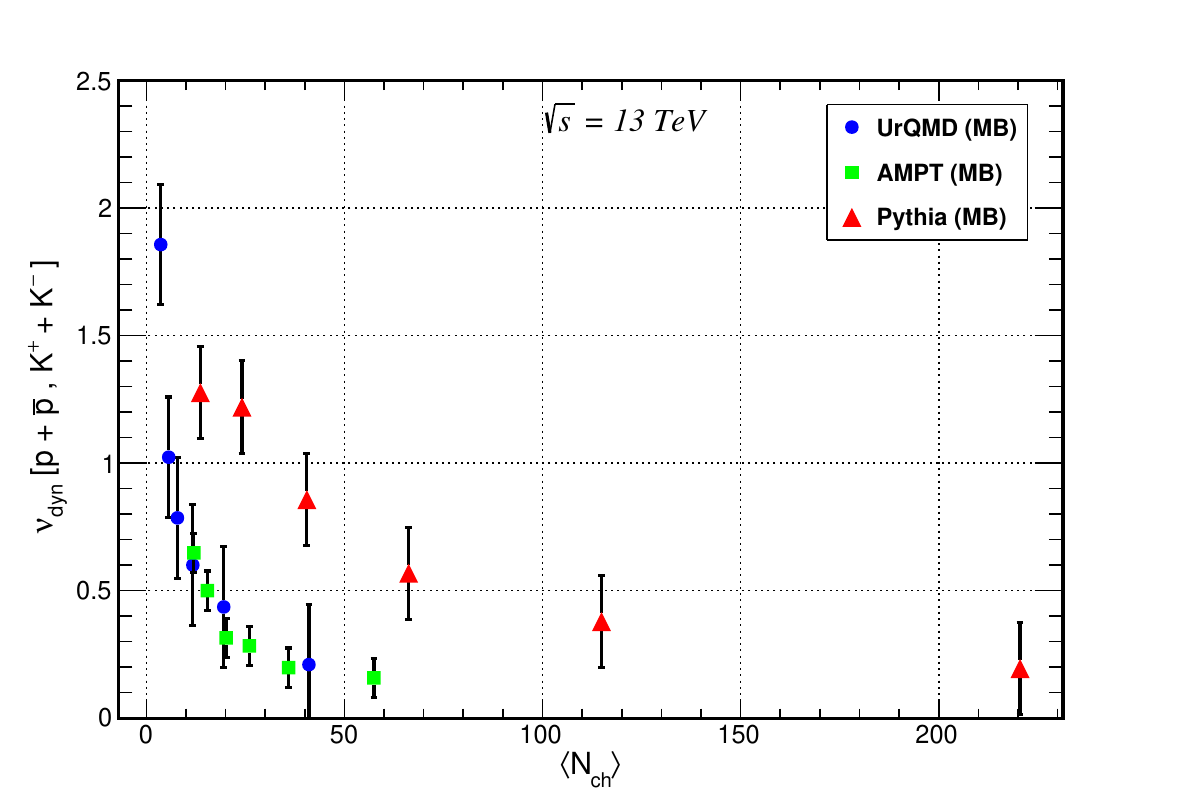}
}
\caption{The variation of $\nu_{dyn}[p, K]$ with mean charged particle multiplicity $\langle N_{ch} \rangle$ at $\sqrt{s} = 13$ TeV.}
\end{figure}

In figs.~$12-15$, the fluctuation between $p$ ($p+\bar{p}$) and $K$ ($K^+ + K^-$) with $\langle N_{ch} \rangle$ is plotted for energies $\sqrt{s}= 2.76, 5.02, 7$ and $13$ TeV respectively. It is clear to note, the value of $\nu_{dyn}[p,K]$ with $\langle N_{ch} \rangle$ decreases non-linearly as $\langle N_{ch} \rangle$ increases for the generated data sets of Pythia, AMPT and UrQMD. At a certain energy, the value of $\nu_{dyn}[p, K]$ is found to be more prominant for UrQMD data and the value of $\nu_{dyn}[p, K]$ is least for AMPT data among all the data sets of Pythia, AMPT and UrQMD.
As the centre of mass energy increases from $\sqrt{s}= 2.76$ TeV to $\sqrt{s}= 13$ TeV, for a given generated data set, the fluctuation in $p$ and $K$ ratio decreases by its magnitude for corresponding value of $\langle N_{ch} \rangle$.

\begin{figure}[h!]
\centering{
\includegraphics[width=.7\columnwidth]{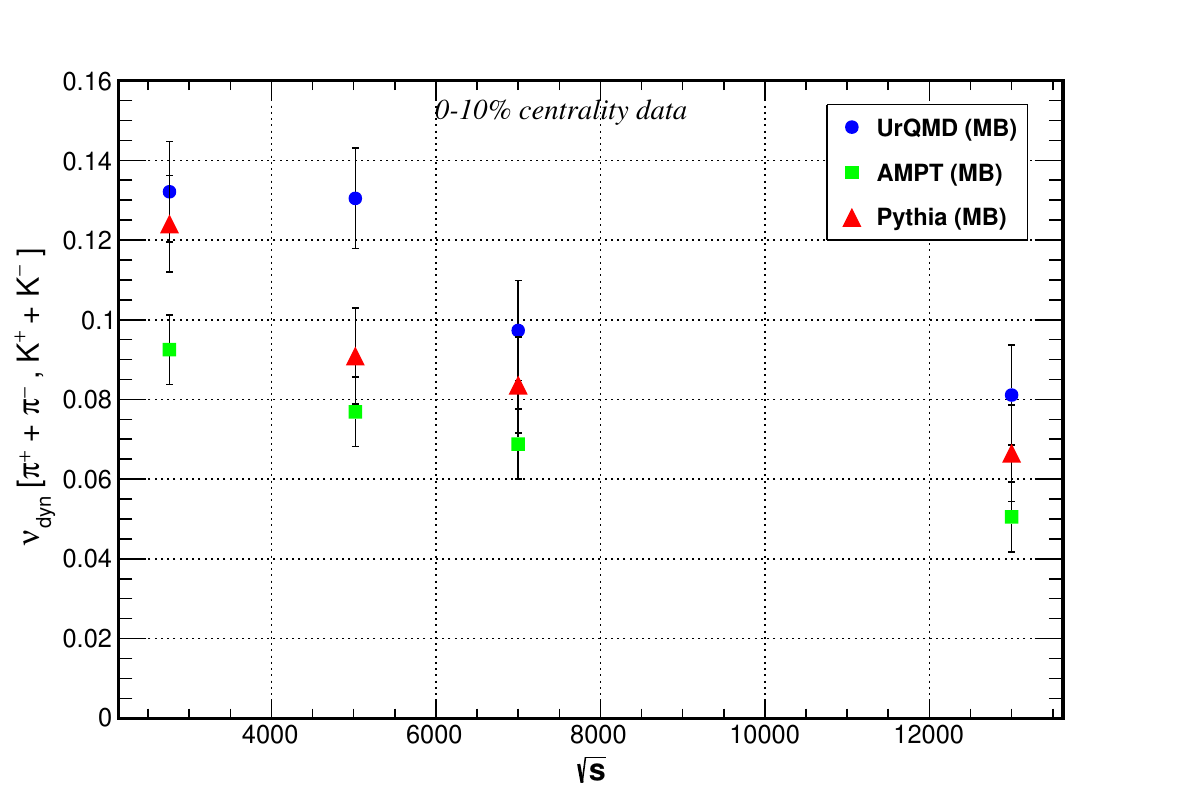}
}
\caption{The variation of $\nu_{dyn}[\pi, K]$ with different c.m. energies $\sqrt{s}$ at $0-10\%$ centrality data.}
\end{figure}

\begin{figure}[h!]
\centering{
\includegraphics[width=.7\columnwidth]{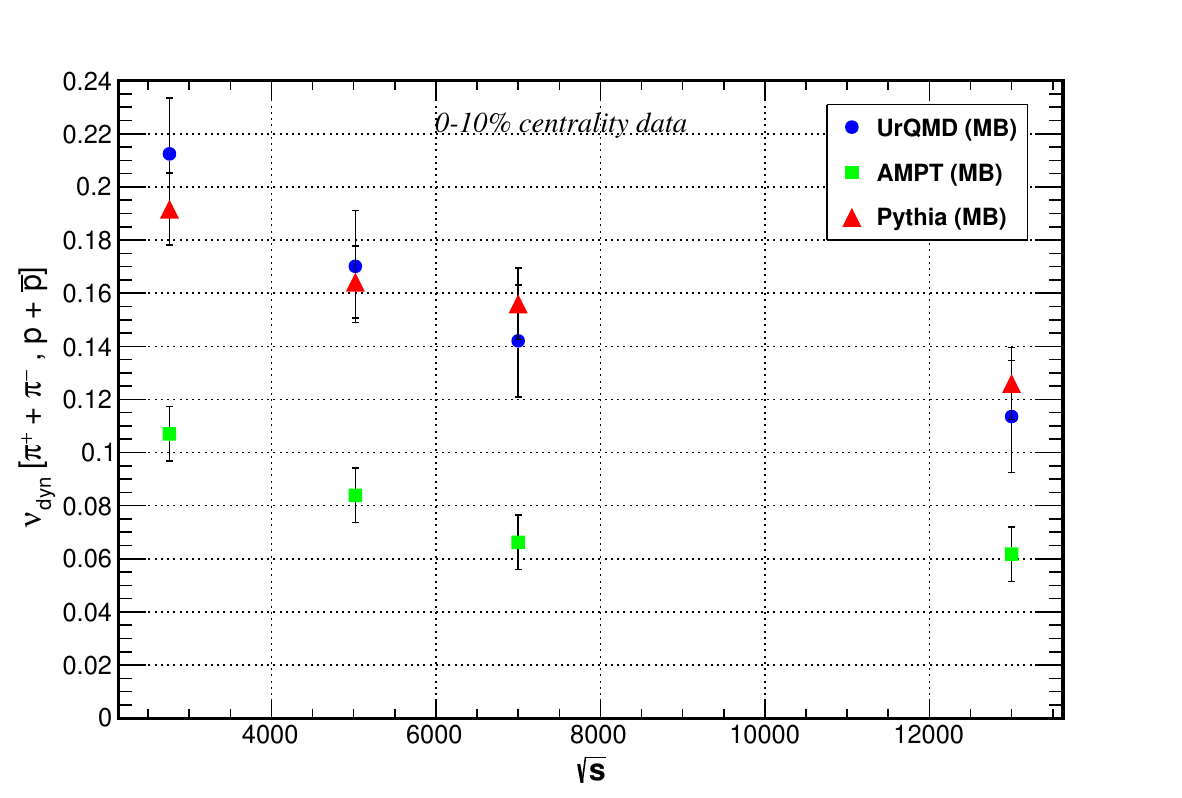}
}
\caption{The variation of $\nu_{dyn}[\pi, p]$ with different c.m. energies $\sqrt{s}$ at $0-10\%$ centrality data.}
\end{figure}

\begin{figure}[h!]
\centering{
\includegraphics[width=.7\columnwidth]{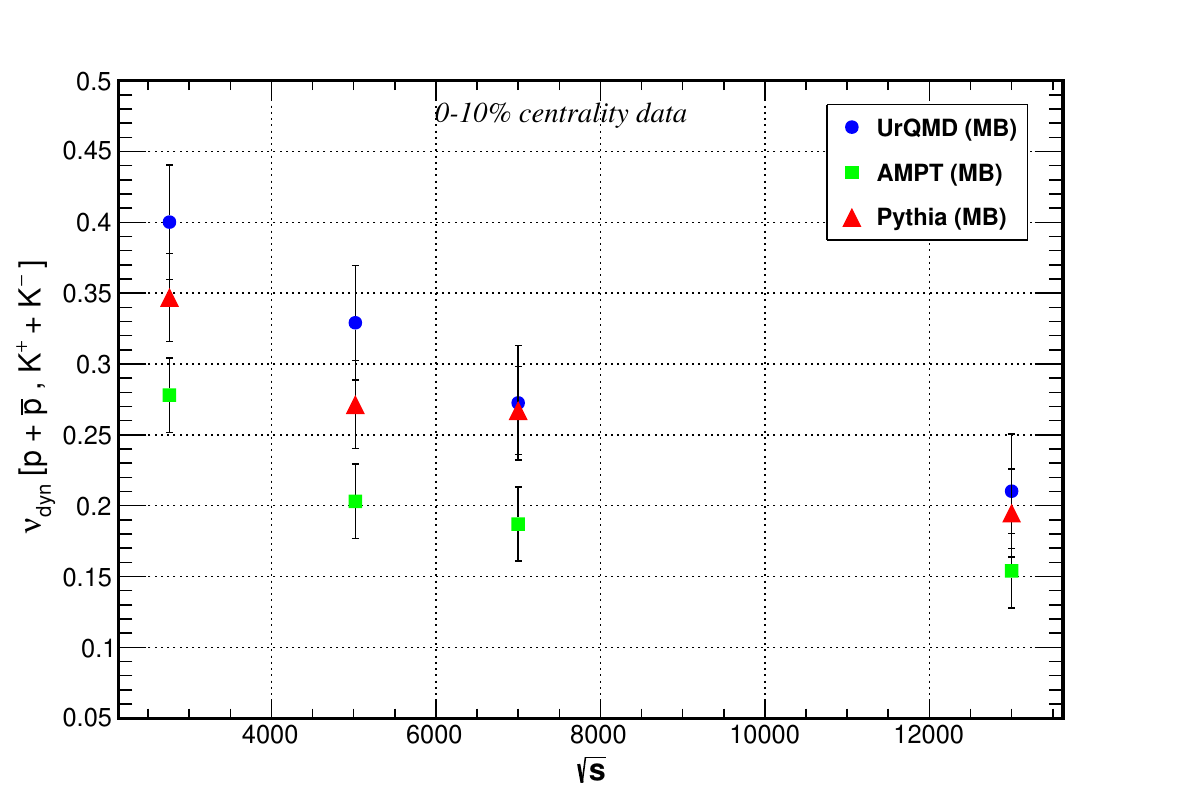}
}
\caption{The variation of $\nu_{dyn}[p, K]$ with different c.m. energies $\sqrt{s}$ at $0-10\%$ centrality data.}
\end{figure}

Figs.~$15-18$ represents the energy-wise variation of $\nu_{dyn}$ of $0-10\%$ centrality data for particle pairs $[\pi, K]$, $[\pi, p]$ and $[p, K]$ respectively. It may be of interest to note, at a certain energy for Pythia and UrQMD data, the magnitude of $\nu_{dyn}$ is least for the particle pair $[\pi, K]$ and the value of $\nu_{dyn}$ is appreciably more for the particle pair $[\pi, p]$ \& $[p, K]$ among all the three particle pairs discussed, which is perhaps due to presence of strong correlation between $\pi$ and $K$ produced in these models over the weaker correlations between $[\pi, p]$ \& $[p, K]$ combination. But the AMPT data shows, the value of $\nu_{dyn}$ is low for the particle pair $[\pi, K]$ \& $[\pi, p]$ than that of the value of $\nu_{dyn}[p, K]$, which indicates that the correlation between $[\pi, K]$ \& $[\pi, p]$ combination is more than the $[p, K]$ combination. 

It is also interesting to note, for a certain particle pair, the fluctuation in that particle pair ratio gradually decreases with increasing LHC energy for all the three models Pythia, AMPT and UrQMD. However, the AMPT data always shows least value of $\nu_{dyn}$ for particle pairs considered, whereas the UrQMD model sometime fails to reproduce a better result towards the discussion of particle ratio fluctuation at LHC energies. The Pythia follows a good trend in producing a generalised pattern for the variation of $\nu_{dyn}$ either with $\langle N_{ch} \rangle$ or with $\sqrt{s}$ for $pp$ collisions at all the discussed LHC energies. 


\section{Conclusion}
The present work is to find the charged particle ratio fluctuations among the conserved charged mesons and baryons like $\pi, K$ and $p$ produced in $pp$ collisions at four different LHC energy, $\sqrt{s}= 2.76, 5.02, 7$ and $13$ TeV using three different simulation models- Pythia, AMPT, and UrQMD. 
 Our effort to study the $\nu_{dyn}$ variable in terms of centrality dependent quantity $\langle N_{ch} \rangle$ and $\sqrt{s}$ for three different particle pairs $[\pi, K], [\pi, p]$ \& $[p, K]$ has lead to the following conclusions:
 
\begin{itemize}
\item[1.] It is observed that for all the three pairs of particle species, at a certain LHC energy, the magnitude of $\nu_{dyn}$ increases on moving from central to peripheral collisions for all the data sets generated by Pythia, AMPT, and UrQMD. The value of $\nu_{dyn}$ for all three pair of particle species is found to be positive, which indicate that the correlation between two particles in a pair is less or the anti-correlation between the two particles in a pair is more. 

\item[2.] In case of Pythia and UrQMD data, at certain energy for most central collisions ($0-10\%$), the little low value of $\nu_{dyn}[\pi, K]$ among the value of $\nu_{dyn}$ of particle pairs $[\pi, K], [\pi, p]$ \& $[p, K]$ is perhaps due to a strong correlation between $\pi$ \& $K$ involved in the generated data of the models and/or the multiplicity distribution of $\pi$ \& $K$ are broader, whereas the higher value of $\nu_{dyn}$ is found for $[\pi, p]$ \& $[p, K]$ combination may be due to weaker $\pi-p$ and $p-K$ correlations present in the Pythia and UrQMD model. For the AMPT model, at a certain energy for most central collisions ($0-10\%$), the value of $\nu_{dyn}$ of $[\pi, K]$ and $[\pi, p]$ is much low compared to the other two models Pythia and UrQMD and the value of $\nu_{dyn}[p, K]$ is little more in its own comparison but still remains very less than the other two models, which indicates that the correlation between $\pi - K$ and $\pi - p$ is more than the correlation between $p - K$. This is perhaps due to the resonance decay process involved in the AMPT model.

\item[3.] The value of $\nu_{dyn}$ for all the three pairs of particle species generated by Pythia, AMPT, and UrQMD is observed to be decreasing with an increase in LHC energy from $\sqrt{s} = 2.76$ TeV to $\sqrt{s} = 13$ TeV for $pp$ collisions, indicate towards the increment in the correlation between particles considered in a pair.  
\end{itemize}

\section*{Acknowledgement}
The author S. Paul expresses gratitude for the financial support granted under  G.O No. 52-Edn(B)/5B-15/2017 dt. 7.6.2017 read with 65-Edn(B)/5-15/2017 dt. 11.7.2017 for Swami Vivekananda Merit-cum-Means Scholarship, Government of West Bengal, India. Similarly, the author T. Biswas extends appreciation for the Inspire fellowship (No. DST/INSPIRE Felloship/2022/IF220173) from the Department of Science and Technology, Govt. of India. 

\section*{Conflict of interests}
The authors declare that there is no conflict of interests regarding the publication of this paper.

\section*{Data availability statement}
In this paper, the simulated data has been generated by using UrQMD, AMPT \& Pythia simulation model. All data generated or analysed during this study are included in this paper. 

\bibliography{references}  


\end{document}